\documentstyle[prl,aps,preprint,eqsecnum,epsf]{revtex}
\tightenlines
\newcommand{\be}{\begin{equation}}
\newcommand{\ee}{\end{equation}}
\newcommand{\ba}{\begin{eqnarray}}
\newcommand{\ea}{\end{eqnarray}}
\begin{document}
\draft
\title{Can Martinez's conjecture be extended to string
theory?}
 \author{Jiliang Jing $^{* \ a\ b}$\footnotetext[1]
 {email: jljing@hunnu.edu.cn} \ \
 \ \ Shiliang Wang$^c$}
\address{ a) Institute of Physics, Hunan
Normal University, Changsha,  Hunan 410081, P. R. China;  \\
b) Department of Astronomy and Applied Physics, University of
Science and Technology of China,\\  Hefei, Anhui 230026, P. R.
China;\\  c) Department of Physics, Central South University,
Changsha, Hunan 410083, P. R. China}

\maketitle
\begin{abstract}

The universality of Martinez's conjecture, which states that the
quasilocal energy of a black hole at the outer horizon reduces to
twice its irreducible mass, or equivalently, to $\sqrt{A/4\pi}$
($A$ is the area of the black hole), is investigated by
calculating Brown-York quasilocal energies for stationary black
holes in heterotic string theory, e. g., for the stationary
Kaluza-Klein black hole, the rotating Cveti$\check{c}$-Youm black
hole, the stationary axisymmetric Einstein-Maxwell-dilaton-axion
black hole, and the Kerr-Sen black hole.  It is shown that
Martinez's conjecture can be extended from general relativity to
heterotic string theory since the quasilocal energies of these
stationary black holes tend to their Arnowitt-Dener-Misner masses
at spatial infinity, and reduce to $\sqrt{A/4\pi}$ at the event
horizons.

\vspace*{1.0cm}


\end {abstract}

 \section{Introduction}

The definition of energy continues to be one of the most active
areas of research in gravitational physics \cite{Bondi}-
\cite{Hayward1}. It is well known that the spacetime metric in
general relativity describes both the background spacetime
structure and the dynamical aspects of the gravitational field,
but no natural way is known to decompose it into its
``background'' and ``dynamical'' components. Thus, the notion of
local energy can not be obtained without a corresponding
decomposition of the spacetime metric. Although a definition of
the energy cannot be found locally, the appropriate quasilocal
definition has been sought by many authors, where the term
``quasilocal'' was introduced by Penrose \cite{Penrose}, referring
to a compact orientable spatial two-surface.

One popular definition for quasilocal energy (QLE) in
gravitational physics was proposed by Brown and York \cite{Brown}
in 1993. One of the appealing features of the definition is that
the QLE can be obtained from the gravitational action and
Hamiltonian in a bounded region. Consider a spacetime $M$
foliated by a spacelike supersurface $\Sigma_t$ defined by
constant $t$, as well as a timelike boundary $B$, a two
dimensional surface $^2B$ is formed by the intersection of
$\Sigma_t$ with $B$. Assuming that $\Sigma_t$ is orthogonal to
$B$, one can then obtain a stress-energy-momentum defined locally
on the three-boundary $B$, with respect to the boundary metric.
The proper energy surface density is defined by the projection of
the stress-tensor on a two-dimensional surface $^2B$ and the total
QLE is defined by integration over the two-dimensional surface
$^2B$.
 \ba
 E=\varepsilon-\varepsilon_{0}=\frac{1}{\kappa}\int_{^2B}
{\rm d}^2x\sqrt{\sigma}(k-k_{0}), \label{1}
 \ea
where $\sigma$ is the determinant of the metric $\sigma_{ij}$ on
$^{2}B$, $k$ is the trace of the extrinsic curvature of $^{2}B$
as embedded in $\Sigma_{t}$,  $k_{0}$ is the value of $k$ in a
flat spacetime, and $\varepsilon_{0}$ is a reference term proposed
in Ref.\cite{Gib} for the normalization of the energy with respect
to a reference spacetime. We choose the reference spacetime to be
flat since all stationary spacetimes that we will study in this
paper are asymptotically flat. The units are chosen so that
$G=C=\hbar=1$ and $\kappa=8\pi$. This expression for the energy
also appears in the study of self-gravitating systems in thermal
equilibrium, where it plays the role of the thermodynamical
energy conjugate to the inverse temperature
\cite{York}-\cite{Bro}.

Martinez \cite{Mar} made use of the definition (\ref{1}) to
calculate the QLE for the stationary Kerr spacetime. By taking the
boundary surface for the Kerr slice as embedded in both the Kerr
slice and flat three-dimensional space, he found that the QLE for
the Kerr black hole possesses the following special properties:
the energy is a positive, monotonically decreasing function of the
boundary surface radius, approaches the Arnowitt-Deser-Misner
(ADM) mass at spatial infinity,  and reduces to twice the
irreducible mass (or $\sqrt{A/4\pi}$, where $A$ is its area) at
the horizon. The last property is known as Martinez's conjecture.

Recently, Bose and Naing\cite{Bose} explored the
(non)universality of Martinez's conjecture by studying the QLE of
the Kerr-Newman black hole and the Kerr-Sen black hole. They
showed that the Kerr-Newman black hole agrees with Martinez's
conjecture, but the Kerr-Sen black hole does not\cite{Bose}. This
result seems to imply that Martinez's conjecture cannot be used
in string theory  since Kerr-Sen black hole is derived from
string theory. Whether the implication is true remains a question
and should be studied carefully.

The main purpose of this paper is to see whether Martinez's
conjecture is applicable to black holes in heterotic string
theory by studying the QLE for some well-known stationary black
holes in string theory.

The  paper is organized  as follows. The QLE's for the stationary
Kaluza-Klein black hole, the rotating Cveti$\check{c}$-Youm black
hole, and the stationary axisymmetric
Einstein-Maxwell-dilaton-axion (EMDA) black hole are calculated by
using the Brown-York definition in Secs. II, III, and IV,
respectively. The QLE for the Kerr-Sen black hole is given in Sec.
IV as a special case of the EMDA black hole. The last section is
devoted to discussion and conclusions. We follow the conventions
of Ref. \cite{Bose}.

\section{QLE for the stationary Kaluza-Klein black hole}

The four dimensional low-energy action obtained from string
theory is \cite{Frolov87}
\begin{equation}\label{action1}
S=\frac{1}{2 \kappa }\int d^4x\sqrt{-g}\left[{\cal R}-2(\nabla
\phi)^2-e^{-2\alpha\phi}F^{2}\right],
\end{equation}
where  $\phi$  is the dilaton scalar field, $F_{ab}$ is the
Maxwell field, and $\alpha$ is a free parameter which governs the
strength of the coupling of the dilaton to the Maxwell field. The
stationary Kaluza-Klein black hole is described by a solution of
the motion equations obtained from Eq. (\ref{action1})  with
$\alpha=\sqrt{3}$, which takes the form \cite{Frolov87}
 \ba
 ds^2&=&-\frac{1-Z}{B}d t^2-\frac{2aZ \sin^2\theta}{B\sqrt{1-v^2}}dt
 d\varphi+\left[B(r^2+a^2)+a^2\sin^2\theta
 \frac{Z}{B}\right] \sin^2\theta d\varphi^2\nonumber \\
 &+&\frac{B\Sigma}{\triangle_r}
 d r^2+B \Sigma d \theta ^2,\label{kkmetric}
 \ea
where
 \ba
 Z=\frac{2mr}{\Sigma},\ \ \ B=\left(1+\frac{v^2
 Z}{1-v^2}\right)^{1/2}, \ \ \Sigma=r^2+a^2\cos^2\theta, \ \
 \triangle_r=r^2+a^2-2mr,
 \ea
and $a$ and $v$ are the rotation parameter and the velocity of the
boost. The dilaton scalar field, and the vector potential of the
stationary Kaluza-Klein black hole can be expressed as
\begin{eqnarray}
\phi=-\frac{\sqrt{3}}{2}\ln B,  \ \
A_t=\frac{v}{2(1-v^2)}\frac{Z}{B^2}, \ \ A_\varphi=-\frac{a v
\sin^2\theta}{2\sqrt{1-v^2}}\frac{Z}{B^2}. \label{kl}
\end{eqnarray}
The area of the event horizon, ADM mass, charge $Q$, and angular
momentum $J$ of the black hole are given by
 \ba
A=\frac{4\pi}{\sqrt{1-v^2}}(r_+^2+a^2), \ \ \ \
M=\left[1+\frac{v^2}{2(1-v^2)}\right]m, \ \  \ \ Q=\frac{m
v}{1-v^2}, \ \ \ \  J=\frac{m a}{\sqrt{1-v^2}}.\label{kkam}
 \ea

 \subsection{Unreferenced QLE for the stationary Kaluza-Klein
black hole}

We now evaluate the unreferenced QLE for the stationary
Kaluza-Klein black hole within a two-surface $^2B$ with a radial
coordinate, $r=r_b$ ($r_b\ge r_{+}$), embedded in a constant
stationary-time hypersurface $\Sigma_{t}$. In the slow rotation
approximation, $a\ll r_b$, the line-element of the two-surface
$^2B$ can be expressed as
 \ba
ds_B^2&\approx&\sqrt{1+\frac{2mv}{(1-v^2)r_b}}
\left\{\left[r_b^2+\frac{mv^2+
(1-v^2)r_b}{2mv^2+(1-v^2)r_b}a^2\cos^2\theta
\right]d\theta^2\right.\cr&&\left.+\left[r_b^2+\frac{mv^2+
(1-v^2)r_b}{2mv^2+(1-v^2)r_b}a^2+\frac{(2-v^2)m}
{2mv^2+(1-v^2)r_b}a^2\sin^2\theta\right]\sin^2\theta
d\varphi^2\right\}. \label{kkrm}
 \ea
Then, by using the definition (\ref{1}) and standard techniques
developed in Refs. \cite{Brown}\cite{Bose}, we get the
unreferenced term
 \ba
\varepsilon&=&-\frac{\sqrt{r_b^2-2mr_b+a^2}}{12(1-v^2)^2r_b^4}
\left[\frac{(1-v^2)r_b}{2mv^2+(1-v^2)r_b}\right]^{\frac{7}{4}}
\Big\{6r_b^2\Big[2mv^2+(1-v^2)r_b\Big]\Big[3mv^2\cr& &+2(1-v^2)
r_b\Big]-a^2\Big[m^2v^2(4+5v^2)+m(4-v^2-3v^4)r_b+
2(1-v^2)^2r_b^2\Big]\Big\},\label{kurfenergy}
 \ea
where the terms of order $ O(a^4/r_b^4)$ and higher have been
neglected.

 \subsection{Intrinsic metric and referenced term}

In order to get the reference term in the QLE (\ref{1}), we should
find a two-dimensional surface isometric to  the metric
(\ref{kkrm}), which is embedded in a flat three-dimensional slice
with the line-element
 \ba
 \label{krm0} {\rm d}s^2={\rm d}R^2+R^2{\rm d}\Theta
^2+R^2\sin^2\Theta{\rm d}\Phi^2,
 \ea
where $R$, $\Theta$ and $\Phi$ are spherical polar coordinates.
Let the two-dimensional surface embedded in the flat slice be
denoted by $R=F(\Theta)$. Thus the line-element on $^{2}B$ can be
expressed as\cite{Bose}
 \ba
 \label{krm1} {\rm
d}s^2=(\dot{R}^2+R^2\dot{\Theta}^2){\rm d}\theta^2+
                R^2\sin^2\Theta{\rm d}\phi^2,
 \ea
where the overdot denotes the derivative with respect to $\theta$.
The requirement that the above line-element be isometric to
 the metric (\ref{kkrm}) for the slow
rotation case indicates that the following  equations should be
satisfied
 \ba
\label{rmrrkk}
\dot{R}^2+R^2\dot{\Theta}^2&=&\sqrt{1+\frac{2mv}{(1-v^2)r_b}}
\left[r_b^2+\frac{mv^2+
(1-v^2)r_b}{2mv^2+(1-v^2)r_b}a^2\cos^2\theta \right], \cr
R^2\sin^2\Theta&=&\sin^2\theta \sqrt{1+\frac{2mv}{(1-v^2)r_b}}
\left[r_b^2+\frac{mv^2+ (1-v^2)r_b}{2mv^2+(1-v^2)r_b}a^2\right.\cr
&&+\left.\frac{(2-v^2)m}
{2mv^2+(1-v^2)r_b}a^2\sin^2\theta\right].
 \ea
By means of the Bose-Naing method \cite{Bose},  from Eq.
(\ref{rmrrkk}) we find
 \ba
F(\Theta)&=&\left(1+\frac{2mv}{(1-v^2)r_b}\right)^{1/4}
\left[r_b^2+\frac{mv^2+
(1-v^2)r_b}{2mv^2+(1-v^2)r_b}a^2\sin^2\Theta\right.\cr
&&-\left.\frac{(2-v^2)m} {2mv^2+(1-v^2)r_b}a^2\cos
2\Theta\right]^{1/2}.
 \ea
Thus, in the slowly rotating approximation the intrinsic metric on
the two-dimensional surface $^2B$ is described by
 \ba
ds^2&\approx& \left[r_b^2+\frac{mv^2+
(1-v^2)r_b}{2mv^2+(1-v^2)r_b}a^2\sin^2\Theta -\frac{(2-v^2)m}
{2mv^2+(1-v^2)r_b}a^2\cos 2\Theta\right]\cr &&
\sqrt{1+\frac{2mv}{(1-v^2)r_b}}(d\Theta^2+\sin^2\Theta d\Phi^2).
 \ea
Working out the extrinsic curvature $k_0$ and then substituting
it into the expression, $\varepsilon_0=\frac{1}{\kappa}
\int_{^2B}k_0\sqrt{\sigma}d\Theta d\Phi$, we obtain the reference
term for the QLE
 \ba
\varepsilon_0&=&-\left(\frac{(1-v^2)r_b}{2mv^2+(1-v^2)r_b}
\right)^{\frac{3}{4}}
\frac{6r_b^2\Big[2mv^2+(1-v^2)r_b\Big]+a^2\Big[(2+v^2)m+
2(1-v^2)r_b\Big]}{6r_b^2(1-v^2)}, \label{krfenergy}
 \ea
where we neglected terms of order $ O(a^4/r_b^4)$ and higher.

 \subsection{Referenced QLE for the
Kaluza-Klein black hole}

Substituting the unreferenced term (\ref{kurfenergy}) and the
referenced term (\ref{krfenergy}) into expression for the QLE
(\ref{1}), we get the referenced QLE for the stationary
Kaluza-Klein black hole
 \ba
E&=& \left(\frac{(1-v^2)r_b}{2mv^2+(1-v^2)r_b}
\right)^{\frac{3}{4}}
\frac{6r_b^2\Big[2mv^2+(1-v^2)r_b\Big]+a^2\Big[(2+v^2)m+
2(1-v^2)r_b\Big]}{6r_b^2(1-v^2)}\cr &&
-\frac{\sqrt{r_b^2-2mr_b+a^2}}{12(1-v^2)^2r_b^4}
\left[\frac{(1-v^2)r_b}{2mv^2+(1-v^2)r_b}\right]^{\frac{7}{4}}
\Big\{6r_b^2\Big[2mv^2+(1-v^2)r_b\Big]\Big[3mv^2\cr& &+2(1-v^2)
r_b\Big]-a^2\Big[m^2v^2(4+5v^2)+m(4-v^2-3v^4)r_b+
2(1-v^2)^2r_b^2\Big]\Big\} .\label{kenergy}
 \ea
In the asymptotic limit $r_b\rightarrow \infty$, we have
 \ba
E(r_b\rightarrow\infty)&=&m\left[1+\frac{v^2}{2(1-v^2)}
\right],\label{kkin}
 \ea
which equals the ADM mass given by Eq. (\ref{kkam}). For all
values of $r_b\geq r_+$, the QLE (\ref{kenergy}) within a constant
radius surface is positive, and is a monotonically decreasing
function of $r_b$. At the event horizon $r_b=r_+$ of the
stationary Kaluza-Klein black hole, the QLE (\ref{kenergy})
becomes
 \ba
E(r_+)&=&r_+\left(\frac{1}{1-v^2}\right)^{\frac{1}{4}}
\left[1+\frac{a^2}{2r_+^2}+
O\left(\frac{a^4}{r_+^4}\right)\right]\cr & \simeq &
\sqrt{\frac{A}{4\pi}},\label{kker}
 \ea
where $A$ is the area of the Kaluza-Klein black hole.

\section{QLE for
Rotating Cveti$\check{c}$-Youm  black hole in string theory}

The  effective action of a toroidally compactified heterotic
string in $D$-dimensions takes the form\cite{Maharana93,Sen94}
 \ba
S&=&\frac{1}{2 \kappa }\int d^Dx \sqrt{-g}\left[{\cal R}-{1\over
(D-2)} g^{\mu\nu}\partial_{\mu}\phi\partial_{\nu}\phi+{1\over 8}
g^{\mu\nu}{\rm Tr}(\partial_{\mu}ML\partial_{\nu}ML)\right.\cr &
&-\left.{1\over{12}}
e^{-2\alpha\phi}g^{\mu\mu^{\prime}}g^{\nu\nu^{\prime}}
g^{\rho\rho^{\prime}}H_{\mu\nu\rho}H_{\mu^{\prime}\nu^{\prime}
\rho^{\prime}}  -{1\over
4}e^{-\alpha\phi}g^{\mu\mu^{\prime}}g^{\nu\nu^{\prime}} {\cal
F}^{i}_{\mu\nu}(LML)_{ij} {\cal
F}^{j}_{\mu^{\prime}\nu^{\prime}}\right], \label{effaction} \ea
 where
$g\equiv {\rm det}\,g_{\mu\nu}$, ${\cal R}_g$ is the Ricci scalar
of $g_{\mu\nu}$, $\phi$ is the dilaton field, ${\cal
F}^i_{\mu\nu} =
\partial_{\mu} {\cal A}^i_{\nu}-\partial_{\nu} {\cal A}^i_{\mu}$
are the $U(1)^{36-2D}$ gauge field strengths,
$H_{\mu\nu\rho}=(\partial_\mu B_{\nu\rho}+2{\cal
A}_{\mu}^{i}L_{ij}{\cal F}^{j}_{\nu\rho})+{\rm cyclic \
permutations\ of} \ \mu, \nu, \rho,$
 and $M$ is the O(10-D,
26-D) symmetric matrix and $L$ is
\begin{equation} L =\left ( \matrix{0 & I_{10-D}& 0\cr I_{10-D}
& 0& 0 \cr 0 & 0 &  I_{26-D}} \right ). \label{4dL}
\end{equation}

The metric of the non-extreme dyonic rotating black hole in terms
of the four-dimensional bosonic fields is of the following form
 \cite{Cvetic96}
 \ba
ds^2&=&\Delta^{1\over 2}\Big\{-{{r^2-2mr+a^2{\rm
\cos}^2\theta}\over \Delta}dt^2+{{dr^2}\over{r^2-2mr+a^2}} +
d\theta^2 +{{{\rm sin}^2\theta}\over \Delta}\Big[f(r)+Wa^2 \cr
&+&a^2(1+{\rm \cos}^2\theta)r^2+2ma^2r{\rm
sin}^2\theta\Big]d\phi^2 -{{4ma}\over \Delta}\Big[({\rm \cosh}
\delta_{p_1}{\rm \cosh} \delta_{p_2}{\rm \cosh} \delta_{e_1}{\rm
\cosh} \delta_{e_2} \cr &-&{\rm \sinh} \delta_{p_1}{\rm \sinh}
\delta_{p_2} {\rm \sinh} \delta_{e_1}{\rm \sinh} \delta_{e_2})r
+2m{\rm \sinh}\delta_{p_1}{\rm \sinh}\delta_{p_2}{\rm
\sinh}\delta_{e_1} {\rm \sinh}\delta_{e_2}\Big]{\rm sin}^2 \theta
dtd\phi\Big\}, \label{4dsol} \ea with
 \ba \Delta &\equiv& f(r) +(2a^2r^2+Wa^2){\rm \cos}^2\theta,
 \cr W&\equiv& 2m q_1   r +4m^2 q_2   +a^2{\rm \cos}^2
\theta,\cr
 f(r)&=&(r+2m{\rm \sinh}^2 \delta_{p_1}) (r+2m{\rm
\sinh}^2 \delta_{p_2})(r+2m{\rm \sinh}^2 \delta_{e_1}) (r+2m{\rm
\sinh}^2 \delta_{e_2}), \cr
 q_1&=& ({\rm \sinh}^2\delta_{p_1}+{\rm
\sinh}^2\delta_{p_2}+ {\rm \sinh}^2\delta_{e_1}+{\rm
\sinh}^2\delta_{e_2}), \cr q_2&=& (2{\rm \cosh}\delta_{p_1}{\rm
\cosh}\delta_{p_2}{\rm \cosh} \delta_{e_1}{\rm
\cosh}\delta_{e_2}{\rm \sinh}\delta_{p_1}{\rm \sinh}
\delta_{p_2}{\rm \sinh}\delta_{e_1}{\rm \sinh}\delta_{e_2}\cr
&-&2{\rm \sinh}^2 \delta_{p_1}{\rm \sinh}^2 \delta_{p_2}{\rm
\sinh}^2 \delta_{e_1}{\rm \sinh}^2 \delta_{e_2}-{\rm \sinh}^2
\delta_{p_2} {\rm \sinh}^2 \delta_{e_1}{\rm \sinh}^2 \delta_{e_2}
\cr &-&{\rm \sinh}^2 \delta_{p_1}{\rm \sinh}^2 \delta_{e_1} {\rm
\sinh}^2 \delta_{e_2}-{\rm \sinh}^2 \delta_{p_1}{\rm \sinh}^2
\delta_{p_2}{\rm \sinh}^2 \delta_{e_2}-{\rm \sinh}^2 \delta_{p_1}
{\rm \sinh}^2 \delta_{p_2}{\rm \sinh}^2 \delta_{e_1}),\cr
 B_{12}&=&-{{2ma{\rm \cos}\theta({\rm
\sinh}\delta_{p_1}{\rm \cosh} \delta_{p_2}{\rm
\cosh}\delta_{e_1}{\rm \sinh}\delta_{e_2}-{\rm \cosh}
\delta_{p_1}{\rm \sinh}\delta_{p_2}{\rm \sinh}\delta_{e_1}{\rm
\cosh} \delta_{e_2})}\over{(r+2m{\rm \sinh}^2
\delta_{p_1})(r+2m{\rm \sinh}^2 \delta_{e_2})+a^2{\rm
\cos}^2\theta}},  \cr e^{\phi}&=&{{(r+2m{\rm \sinh}^2
\delta_{p_1})(r+2m{\rm \sinh}^2 \delta_{p_2})+a^2{\rm \cos}^2
\theta}\over \Delta^{1\over 2}},
 \label{4ddef} \ea
 where $\delta_{e_1}$,  $\delta_{e_2}$, $\delta_{p_1}$, and
$\delta_{p_2}$ are four boosts, and $a$ and $m$ represent the
rotational and mass parameters, respectively.

The ADM mass, $U(1)$ charges $Q^{(1),(2)}_2, P^{(1),(2)}_1$, and
 angular momentum $J$ of the solution can be expressed as  \ba
M&=&\frac{1}{2}m({\rm \cosh}^2 \delta_{e_1}+{\rm \cosh}^2
\delta_{e_2}+ {\rm \cosh}^2 \delta_{p_1}+{\rm \cosh}^2
\delta_{p_2})-m, \cr Q^{(1)}_2 &=&\frac{1}{2}m{\rm
\cosh}\delta_{e_1}{\rm \sinh}\delta_{e_1},\ \ \ \ \ Q^{(2)}_2 =
\frac{1}{2}m{\rm \cosh}\delta_{e_2}{\rm \sinh}\delta_{e_2}, \cr
P^{(1)}_1 &=&\frac{1}{2}m{\rm \cosh}\delta_{p_1}{\rm
\sinh}\delta_{p_1},\ \ \ \ \ P^{(2)}_1 = \frac{1}{2}m{\rm
\cosh}\delta_{p_2}{\rm \sinh}\delta_{p_2}, \cr J&=&ma({\rm
\cosh}\delta_{e_1}{\rm \cosh}\delta_{e_2}{\rm \cosh}\delta_{p_1}
{\rm \cosh}\delta_{p_2}-{\rm \sinh}\delta_{e_1}{\rm
\sinh}\delta_{e_2} {\rm \sinh}\delta_{p_1}{\rm
\sinh}\delta_{p_2}). \label{4dphys}
 \ea
The black hole (\ref{4dsol}) has the inner $r_-$ and the outer
$r_+$ horizons at
\begin{equation}
r_{\pm}=m\pm \sqrt{m^2-a^2}, \label{horizon}
\end{equation}
provided $m\ge a$. The area of the event horizon is given by
 \ba A&=&4\pi\Big[
(r_++2m\sinh^2 \delta_{e_1})(r_++2m\sinh^2
\delta_{e_2})(r_++2m\sinh^2 \delta_{p_1})(r_++2m\sinh^2
\delta_{p_2})\cr &&+2a^2(2m^2q_2+mq_1r_++r_+^2) +a^4\Big]^{1/2}.
\label{cya} \ea

\subsection{Unreferenced QLE for the Rotating
Cveti$\check{c}$-Youm  black hole}

Using the definition of the unreferenced QLE for the rotating
Cveti$\check{c}$-Youm  black hole within a surface $^2B$ with
radial coordinate $r=r_b$, we have
 \ba
\varepsilon&=&-\int_0^\pi\frac{\sqrt{r_b^2-2mr_b+a^2}} {8
\sqrt{g_{\varphi \varphi }}\Delta^{1/2}} \left[\left(\frac{1}
{r_b+2m\sinh^2\delta_{e_1}}+\frac{1} {r_b+2m\sinh^2\delta_{e_2}}+
\frac{1} {r_b+2m\sinh^2\delta_{p_1}}\right.\right.\cr
&&\left.\left.+ \frac{1} {r_b+2m\sinh^2\delta_{p_2}}\right)f(r_b)+
2ma^2(q_1+\sin^2\theta)+r_b(3+\cos2\theta)a^2\right] \sin^2\theta
d\theta.
 \ea
Obviously, the integral cannot be expressed in simple functions.
But in the slowly rotating approximation, i.e.,  $a/r_{b}\ll 1$,
the above integral approximately becomes
 \ba
\varepsilon&=&-\frac{\sqrt{r_b^2-2mr_b+a^2}}{12 f(r_b)
^{7/4}}\Big\{6f(r_b)\left[ 4m^3\left(
\sinh^2\delta_{e_2}\sinh^2\delta_{p_1}\sinh^2\delta_{p_2}
 +\sinh^2\delta_{e_1}\left(
\sinh^2\delta_{p_1}\right.\right.\right.\cr &\times
&\left.\left.\left. \sinh^2\delta_{p_2}+\sinh^2\delta_{e_2}\left(
\sinh^2\delta_{p_1} +\sinh^2\delta_{p_2} \right)  \right) \right)
+4m^2r_b\left(\sinh^2\delta_{p_1}\sinh^2\delta_{p_2} +
\sinh^2\delta_{e_2}\right. \right.\cr &\times &\left.\left.(
\sinh^2\delta_{p_1}+\sinh^2\delta_{p_2}) +\sinh^2\delta_{e_1}(
\sinh^2\delta_{e_2} + \sinh^2\delta_{p_1} + \sinh^2\delta_{p_2})
\right) + 3m\left( \sinh^2\delta_{e_1} \right.\right.\cr &+
&\left.\left. \sinh^2\delta_{e_2}+ \sinh^2\delta_{p_1} +
\sinh^2\delta_{p_2} \right) r_b^2 + 2r_b^3 \right]+ \left[
8m^5\left(
-7\sinh^2\delta_{e_2}\sinh^2\delta_{p_1}\sinh^2\delta_{p_2} q_2
\right.\right.\cr &+ &\left.\left.
\sinh^2\delta_{e_1}\left(-7\sinh^2\delta_{p_1}
\sinh^2\delta_{p_2}q_2 + \sinh^2\delta_{e_2}\left(
\sinh^2\delta_{p_1}\left( 4\sinh^2\delta_{p_2}\left( 2 + 3q_1
\right)  - 7q_2 \right) \right.\right.\right.\right.\cr &-
&\left.\left.\left.\left. 7\sinh^2\delta_{p_2}q_2 \right) \right)
\right)  + 4m^4\left( -14\sinh^2\delta_{p_1}\sinh^2\delta_{p_2}q_2
+\sinh^2\delta_{e_2}\left( \sinh^2\delta_{p_1} \left(
\sinh^2\delta_{p_2}\left( 4 \right.\right.\right.\right.\right.\cr
&+ &\left.\left.\left.\left. \left. 5q_1 \right)  - 14q_2 \right)
- 14\sinh^2\delta_{p_2}q_2 \right)+ \sinh^2\delta_{e_1}\left(
\sinh^2\delta_{p_1}
 \left( \sinh^2\delta_{p_2}\left( 4 + 5q_1 \right)  - 14 q_2 \right)
 \right.\right.\right.\cr &+ &\left.\left.\left.
  \sinh^2\delta_{e_2}\left( \sinh^2\delta_{p_2}\left( 4 + 5q_1
  \right)  +
\sinh^2\delta_{p_1}\left( 4 + 32\sinh^2\delta_{p_2} + 5q_1 \right)
 -14q_2 \right)\right.\right.\right.\cr &- &\left.\left.\left.
 14\sinh^2\delta_{p_2}q_2
\right)  \right)r_b + 2m^3\left(
-2\sinh^2\delta_{p_1}\sinh^2\delta_{p_2}q_1
+\sinh^2\delta_{e_2}\left(
\sinh^2\delta_{p_1}\left(22\sinh^2\delta_{p_2}
\right.\right.\right.\right.\cr &- &\left.\left.\left.\left. 2q_1
\right)-
 2\sinh^2\delta_{p_2}q_1 - 21q_2 \right)  +
 \sinh^2\delta_{e_1}\left( \sinh^2\delta_{p_1}
 \left( 22\sinh^2\delta_{p_2} - 2q_1 \right) \right.\right.\right.\cr
&+&\left.\left.\left.\sinh^2\delta_{e_2}\left(
22\sinh^2\delta_{p_1} + 22\sinh^2\delta_{p_2} - 2q_1 \right) -
2\sinh^2\delta_{p_2}q_1 - 21q_2 \right) - 21\sinh^2\delta_{p_1}q_2
\right.\right.\cr &- &\left.\left. 21\sinh^2\delta_{p_2}q_2
\right) r_b^2 + m^2\left( -4\sinh^2\delta_{p_1} -
4\sinh^2\delta_{p_2} + 12\sinh^2\delta_{p_1}\sinh^2\delta_{p_2} +
\sinh^2\delta_{e_2} \right.\right.\cr &\times &\left.\left. \left(
-4 + 12\sinh^2\delta_{p_1} + 12\sinh^2\delta_{p_2} - 9q_1 \right)
+ \sinh^2\delta_{e_1}\left( -4 + 12\sinh^2\delta_{e_2} +
12\sinh^2\delta_{p_1} \right.\right.\right.\cr &+
&\left.\left.\left. 12\sinh^2\delta_{p_2} - 9q_1 \right)  -
9\sinh^2\delta_{p_1}q_1 -9\sinh^2\delta_{p_2}q_1 - 28q_2 \right)
r_b^3 + m\left( -4 + \sinh^2\delta_{e_1} \right.\right.\cr &+
&\left.\left. \sinh^2\delta_{e_2} + \sinh^2\delta_{p_1} +
\sinh^2\delta_{p_2} - 8q_1 \right) r_b^4 - 2r_b^5
\right]a^2\Big\}+r_bO\left(\frac{a^4}{r_b^4}\right),\label{urfenergy}
 \ea
which is the QLE without a reference term for the rotating
Cveti$\check{c}$-Youm  black hole.

 \subsection{Intrinsic metric and referenced term}

To get the reference term in the QLE for the rotating
Cveti$\check{c}$-Youm  black hole, we should find a
two-dimensional surface isometric to the $\theta-\varphi$ surface
of the metric (\ref{4dsol}), which is embedded in a flat
three-dimensional slice  with the line element (\ref{krm0}).
following the discussion of the preceding section, we know that
the requirement that the line element be isometric to the
$\theta-\varphi$ surface of the metric (\ref{4dsol}) for slow
rotation case indicates that the following  equations should be
satisfied:
 \ba
\label{rmrr} \dot{R}^2+R^2\dot{\Theta}^2&=&\sqrt{f(r_b)}\left
(1+\frac{W+2r^2}{2f(r_b)}a^2\cos^2\theta \right ),\cr
R^2\sin^2\Theta&=&\frac{\sin^2\theta}{\sqrt{f(r_b)}}
\left[f(r_b)+(W+2r^2)a^2\left(1-\frac{1}{2}\cos^2\theta\right)
+(2mr-r^2)a^2\sin^2\theta\right].
 \ea
Solving the equations, we get
 \ba
F(\Theta)=\frac{1}{f(r_b)^{1/4}}\left[f(r_b)+(W+2r^2)
a^2\left(\sin^2\Theta-\frac{1}{2}\cos^2\Theta\right)
+(r^2-2mr)a^2\cos2\Theta\right]^{\frac{1}{2}},
 \ea
where $F(\Theta)=R$ denotes the two-dimensional surface embedded
in the flat slice. Then, in the slowing rotating approximation the
intrinsic metric on the two-dimensional surface $^2B$ is given by
 \ba
ds_r^2&\approx& \frac{1}{\sqrt{f(r_b)}}\left[f(r_b)+(W+2r^2)
a^2\left(\sin^2\Theta-\frac{1}{2}\cos^2\Theta\right)
+(r^2-2mr)a^2\cos2\Theta\right]\cr &
&\times(d\Theta^2+\sin^2\Theta d\Phi^2).
 \ea
Working out the extrinsic curvature $k_0$ and then substituting
it into the expression, $\varepsilon_0=\frac{1}{\kappa}
\int_{^2B}k_0\sqrt{\sigma}d\Theta d\Phi$, we obtain reference
term of the QLE:
 \ba
\varepsilon_0&=&-\int_0^\pi\frac{d\Theta \sin\Theta} {2
f(r_b)^{1/4}}\sqrt{f(r_b)+\Big[(W+2r_b^2)\Big(1-
\frac{3\cos^2\Theta}{2}\Big)+(r_b^2-2mr_b)\cos2\Theta\Big]a^2}\cr
 &=&
-\frac{1}{6 f(r_b)^{3/4}}[6f(r_b)+a^2
(6m^2q_2+m(2+3q_1)r_b+2r_b^2)] +r_bO(\frac{a^4}{r_b^4}).
\label{rfenergy}
 \ea

 \subsection{Referenced QLE for Rotating
Cveti$\check{c}$-Youm  black hole}

Substituting the unreferenced term (\ref{urfenergy}) and the
referenced term (\ref{rfenergy}) into the referenced QLE
 \ba
E=\varepsilon-\varepsilon_0,\label{energy}
 \ea
and taking the asymptotic limit $r_b\rightarrow \infty$, we have
 \ba
E(r_b\rightarrow\infty)=\frac{1}{2}m
(\cosh^2\delta_{e_1}+\cosh^2\delta_{e_2}+\cosh^2\delta_{p_1}
+\cosh^2\delta_{p_2} )-m,\label{cyin}
 \ea
which equals the ADM mass given by Eq. (\ref{4dphys}). For all
values of $r_b\geq r_+$, the QLE (\ref{energy}) within a constant
radius surface is positive, and a monotonically decreasing
function of $r_b$. At the event horizon $r_b=r_+$ of the
Cveti$\check{c}$-Youm black hole, the QLE (\ref{energy}) becomes
 \ba
E&=&\frac{[(r_++2m\sinh^2\delta_{e_1})
(r_++2m\sinh^2\delta_{e_2})(r_++2m\sinh^2\delta_{p_1})
(r_++2m\sinh^2\delta_{p_2})]^{\frac{1}{4}}}{2}\cr & &
\left[2+\frac{a^2(2m^2q_2+mq_1r_+
+r_+^2)}{(r_++2m\sinh^2\delta_{e_1})
(r_++2m\sinh^2\delta_{e_2})(r_++2m\sinh^2\delta_{p_1})
(r_++2m\sinh^2\delta_{p_2})}\right]\cr&&+r_+O(\frac{a^4}{r_+^4})\cr
&&\simeq \sqrt{\frac{A}{4\pi}},\label{cyer}
 \ea
where $A$ represents the area of the Cveti$\check{c}$-Youm black
hole, which is defined by Eq. (\ref{cya}).

\section{QLE for the Stationary axisymmetric EMDA black hole}

The four-dimensional low-energy action obtained from heterotic
string theory is \cite{Garcia}
 \ba
 \label{2}
S&=&\frac{1}{2 \kappa}\int{\rm d}^{4}x \sqrt{-{\rm g}}({\cal
R}-2{\rm
g}^{\mu\nu}\bigtriangledown_\mu\Phi\bigtriangledown_\nu\Phi
-\frac{1}{2}{\rm e}^{4\Phi}{\rm
g}^{\mu\nu}\bigtriangledown_\mu{\rm K}_{\rm a}\bigtriangledown_\nu
 {\rm K}_{a} \nonumber \\
& &-{\rm e}^{-2\Phi}{\rm g}^{\mu\lambda}{\rm g}^{\nu\rho}{\rm
F}_{\mu\nu}{\rm F}_{\lambda\rho}-{\rm K}_{a}{\rm
F}_{\mu\nu}\tilde{{\rm F}}^{\mu\nu}),  \ea
 with $ \tilde
{F}_{\mu\nu}=-\frac{1}{2} \sqrt{-{\rm g}}
\varepsilon_{\mu\nu\alpha\beta} F^{\alpha\beta} $,  where $\Phi$
is the massless dilaton field, $ F_{\mu\nu}$ is the
electromagnetic antisymmetric tensor field, and $K_{\rm a}$ is
the axion field dual to the three-index antisymmetric tensor
field  $ H=-\exp(4\Phi)*{\rm d}K_{\rm a}/4 $.

The stationary axisymmetric EMDA black hole solution (we take the
solution $ b=0$ of Eq.(14) in Ref.\cite{Garcia}; the reason we use
this solution is that the solution $b\not=0$ cannot be interpreted
properly as a black hole) is given \cite{Garcia} by \ba
 \label{emdam}
{\rm d}s^2&=&-\frac{\Sigma-a^2\sin^2\theta}{\Delta}{\rm
d}t^2-\frac{2a\sin^2\theta}{\Delta}\left [ (r^2+a^2-2Dr)-\Sigma
\right ] {\rm d}t{\rm d}\phi \nonumber \\&
&+\frac{\Delta}{\Sigma}{\rm d}r^2+\Delta {\rm
d}{\theta}^2+\frac{\sin^2\theta}{\Delta}
\left[(r^2+a^2-2Dr)^2-\Sigma a^2\sin^2\theta\right ] {\rm
d}\phi^{2},  \ea    with \ba \Sigma&=&r^2-2mr+a^2,\ \ \ \
\Delta=r^2-2Dr+a^2\cos^2\theta, \cr  A_t&=&\frac{1}{\Delta}(Q
r-{\rm g}a\cos\theta), \ \ \ \ A_r=A_\theta=0,\ \ \ \
A_{\phi}=\frac{1}{a\Delta}(-Qra^2\sin^2\theta+{\rm
g}(r^2+a^2)a\cos\theta), \cr  {\rm
e}^{2\Phi}&=&\frac{W}{\Delta}=\frac{\omega}{\Delta}
(r^2+a^2\cos^2\theta), \ \ \  \omega={\rm e}^{2\Phi_{0}},\ \ \ \
K_a=K_0+\frac{2aD\cos\theta}{W}. \label{emdafa}
 \ea
 The ADM mass $M$ and the area of the
black hole are given by
 \ba   \label{emdaam} M=m-D,\ \ \ \ \ \  A=4\pi
(r_+^2-2Dr_++a^2),
 \ea
where the horizons $r_{\pm}=(M-(Q^2/2\omega
M))\pm\sqrt{(M-(Q^2/2\omega M))^2-a^2}$.

\subsection{Unreferenced QLE for stationary axisymmetric EMDA black
hole}

It is easily to shown that the line-element on the two-surface
$^2B$ in the slow-rotation approximation (i.e. $a\ll r_b$) can be
expressed as
 \ba \label{10}
{\rm d}s^2&=&r_{B}^{2}\left (
1-\frac{2D}{r_{B}}+\frac{a^2}{r_{B}^{2}}\cos^2\theta\right ){\rm
d}\theta^2 \nonumber\\ & +&r_{B}^{2}\left
\{1-\frac{2D}{r_b}+\frac{a^2}{r_b^2} \left [1+\frac{2(m-D)}{
r_{B}(1-\frac{2D}{r_{B}}) } \sin^2\theta\right ]\right
\}\sin^2\theta {\rm d}\phi ^2,
 \ea
where terms of order $ O(a^4/r_b^4)$ and higher have been
neglected.

By using the expression (\ref{1}) we find that  the unreferenced
term of the QLE for the axisymmetric EMDA black hole within the
surface $^2B$ embedded in $\Sigma_{t}$ is given by
 \ba \label{13}
\varepsilon&=&-\frac{r_b}{2}\sqrt{ 1-\frac{2m}{r_b}
+\frac{a^2}{r_b^{2}} } \int_{0}^{\pi}{\rm d}\theta \sin\theta
\frac{1}{\sqrt{1-\frac{2D}{r_b}+\frac{a^2}{r_b^2} \cos^2\theta }}
\nonumber\\ & &\times \frac{
(1-\frac{D}{r_b})(1-\frac{2D}{r_b}+\frac{a^2}{r_b^2})
 -\frac{a^2}{2r_b}(1-\frac{M}{r_b})\sin^2\theta
 }{\sqrt{(1-\frac{2D}{r_b})^2+\frac{a^2}{r_b^2}
 \left [1+\cos^2\theta-\frac{4D}{r_b}+(\frac{2m}{r_b}
 +\frac{a^2}{r_b^2})\sin^2\theta \right ]}}.\cr
&=&-r_b\sqrt{ 1-\frac{2m}{r_b}+\frac{a^2}{r_b^{2}}}\left
\{\frac{1-\frac{D}{r_b}}{\sqrt{1-\frac{2D}{r_b}}}-
\frac{a^2}{6r_b^2}\left [\frac{1}{\left(1-\frac{2D}
{r_b}\right)^{3/2}}
+\frac{2m-D}{r_b\left(1-\frac{2D}{r_b}\right)^{5/2}} \right
]\right\},
 \ea
where the terms of order $ O(a^4/r_b^4)$ and higher have been
neglected.

\subsection{Intrinsic metric and referenced term}

To calculate the reference term of the QLE for the stationary
axisymmetric EMDA black hole, we should find a two-dimensional
surface isometric to (\ref{10}), which is embedded in a flat
three-dimensional slice with the line-element (\ref{krm0}). The
requirement that the line-element be isometric to (\ref{10})
indicates that
 \ba
   \label{17} \dot{R}^2+R^2\dot{\Theta}^2&=&r_b^2\left
(1-\frac{2D}{r_b}+\frac{a^2}{r_b^2} \cos^2\theta \right ),\cr
  R^2\sin^2\Theta&=&r_b^2\sin^2\theta
 \left \{\left (1-\frac{2D}{r_b}\right )+\frac{a^2}{r_b^2}
  \left [1+\frac{2(m-D)\sin^2\theta}{r_b(1-\frac{2D}{r_b})}
  \right]\right \}.
   \ea
After some calculations we get
 \ba \label{21}
F(\Theta)&=&r_b\left
[\sqrt{1-\frac{2D}{r_b}}+\frac{a^2\sin^2\Theta}
{2r_b^2\sqrt{1-\frac{2D}{r_b}}} -\frac{a^2\cos
2\Theta(m/r_b-\frac{D}{r_b})} {r_b^2(1-\frac{2D}{r_b})^{3/2}}
\right ].
 \ea
This expression implies that the two-surface $R=F(\Theta)$
describes a distorted sphere. Thus, in the slow-rotation
approximation we get the intrinsic metric
 \ba \label{22}
 {\rm d}s^2&\simeq& r_b^2\left
[(1-2D/r_b)+\frac{a^2}{r_b^2}
 \left (\sin^2\Theta-\frac{2(m-D)}{r_b
 (1-2D/r_b)}\cos2\Theta \right )
 \right ] ({\rm d}\Theta^2+\sin^2\Theta{\rm d}\Phi^2 ).
 \ea
The extrinsic curvature $k_{0}$  and its proper integral can be
calculated using the metric. Then the reference term in (\ref{1})
is given by
 \ba   \label{23}
\varepsilon_{0} &=& \frac{1}{\kappa}\int_{^2B} k_{0}
\sqrt{\sigma}{\rm d}\Theta{\rm d}\Phi   \nonumber\\
&\simeq& -r_b \left
\{\sqrt{1-\frac{2D}{r_b}}+\frac{a^2}{3r_b^2\sqrt
{1-\frac{2D}{r_b}}} \left [1+\frac{m-D}{r_b(1-2D/r_b)} \right
]\right \},
 \ea
where terms of order $ O(a^4/r_b^4)$ and higher have been
neglected.

 \subsection{Referenced QLE for the stationary
axisymmetric EMDA black hole}

Substituting Eqs. (\ref{23}) and (\ref{13}) into Eq. (\ref{1}), we
get the reference QLE for the stationary axisymmetric EMDA black
hole
 \ba \label{24}
E &=& r_b \left \{ \sqrt{1-\frac{2D}{r_b}}
 -\frac{1-\frac{D}{r_b}} {\sqrt{1-\frac{2D}{r_b}}}
  \sqrt{ 1-\frac{2m}{r_b} +\frac{a^2}{r_b^2} }\right \}
  +\frac{a^2}{6r_b\sqrt{1-2D/r_b}}     \nonumber\\
  & & \times \left \{ 2+\frac{2(m-D)}{r_b(1-2D/r_b)}+
  \left [\frac{1}{1-\frac{2D}{r_b}}+\frac{2m-D}
   {r_b(1-\frac{2D}{r_b})^2}\right]
    \sqrt{ 1-\frac{2m}{r_b} +\frac{a^2}{r_b^2} }
   \right \}    \nonumber\\
  & & +r_b O(\frac{a^4}{r_b^4}).
 \ea
As $ r_b \rightarrow \infty$, we have
 \ba \label{25}
  E(r_b\rightarrow \infty)= m-D=M,\label{emdain}
 \ea
which is the ADM mass (\ref{emdaam}) of the  stationary
axisymmetric EMDA black hole. For all values of $r_b>r_+$, the
QLE (\ref{24}) within the constant radius surface is positive,
and can be expressed as a monotonically decreasing function of
$r_b$. At the outer horizon $r_b=r_{+}$, the energy (\ref{24})
becomes
 \ba
\label{29} E(r_b=r_{+})&=&r_{+}\left
(\sqrt{1-\frac{2D}{r_{+}}}+\frac{1}{2\sqrt {1-\frac{2D}{r_{+}}}}
\frac{a^2}{r_{+}^2}+ O(\frac{a^4}{r_+^4}) \right )\cr
&\simeq&\sqrt{\frac{A}{4\pi}},\label{emdaer}
 \ea
where the area of the black hole, $A$, is given by Eq.
(\ref{emdaam}).

\subsection{QLE for the Kerr-Sen black hole}

By setting $\omega=1$ and $D=-m\sinh^2(\alpha/2)$, the metric
(\ref{emdam}) becomes the Kerr-Sen black hole \cite{Sen} in string
gravity. The ADM mass, angular momentum $J$, and electric charge
$Q$ of the Kerr-Sen black hole are
  \ba
\label{9} 2M &=& m(1+\cosh\alpha),\;\;
     2J=ma(1+\cosh\alpha) \nonumber\\
\sqrt{2}Q &=& m\sinh\alpha, \;\;\;\;\;
 \sqrt{2}\mu=ma\sinh\alpha,
  \ea
see Eq.(16) and Eq.(17) of Ref.\cite{Sen}.

Thus, from Eq.  (\ref{24}) we get the QLE for the Kerr-Sen black
hole
 \ba   \label{28} E&=&r_b
\sqrt{\lambda}\left [1-\frac{r_b+m\sinh^2\frac{\alpha}{2}}{r_b
\lambda} \sqrt{ 1-\frac{2m}{r_b} +\frac{a^2}{r_b^2} }
\right ] \nonumber \\
 & &+\frac{a^2}{6r_b \sqrt{\lambda}}
\left \{   2+\frac{2m\cosh^2\frac{\alpha}{2}}{r_b \lambda }
+\left [1+\frac{m(1+\cosh^2\frac{\alpha}{2})}{r_b\lambda}
 \right ]\sqrt{ 1-\frac{2m}{r_b} +\frac{a^2}{r_b^2} }
 \right \} ,
 \ea
where $\lambda=1+2m\sinh^2(\alpha/2)/r_b$. We note that the
expression differs from that  in Ref.\cite{Bose}. However,
putting $r_b=r_+$ into Eq. (\ref{28}), we get
 \ba   \label{32}
E_{S}(r_b=r_{+}) =r_{+}\cosh\frac{\alpha}{2} \left
[1+\frac{a^2}{2r_{+}^2}+ O(a^4/r_+^4) \right ]\simeq\sqrt
{\frac{A}{4\pi}}.\label{sener}
 \ea
The expression (\ref{32}) shows that Martinez's conjecture helds
for the Kerr-Sen black hole. This conclusion differs from that of
Bose-Naing \cite{Bose}.

\section{Discussion and Conclusion}

The quasilocal energies for stationary black holes in heterotic
string theory, e. g., for the stationary Kaluza-Klein black hole,
the rotating Cveti$\check{c}$-Youm black hole, the stationary
axisymmetric EMDA black hole, and the Kerr-Sen black hole, are
calculated using the Brown-York approach within two-surfaces of
constant radii embedded in constant stationary-time slices.
Several universal properties of the quasilocal energies are
determined: a) In the asymptotic limit $r_b\rightarrow \infty$,
the quasilocal energies of the stationary black holes tend to
their corresponding ADM masses (see Eqs. (\ref{kkin}),
(\ref{cyin}), and (\ref{emdain})). b) For all values of $r_b\geq
r_+$, the quasilocal energies within a constant radius $r_b$
surface are positive, and are monotonically decreasing functions
of $r_b$. c)   At the event horizon $r_b=r_+$, the quasilocal
energies reduce to $\sqrt{A/4\pi}$ (see Eqs. (\ref{kker}),
(\ref{cyer}), (\ref{emdaer}), and (\ref{sener})). The last
property shows that the Martinez's conjecture, originally
proposed for the Kerr black hole and true for black holes in
general relativity\cite{Mar}\cite{Bose}, can be extended to black
holes in heterotic string theory. The conclusion is different
from that of Bose and Naing \cite{Bose} that Martinez's
conjecture does not hold for the Kerr-Sen black hole in string
theory.

\begin{acknowledgements}
This work was supported by the National Natural Science Foundation
of China under Grant No. 19975018,  and the Theoretical Physics
Special Foundation of China under Grant No. 19947004.
\end{acknowledgements}

\end{document}